\documentclass[conference]{IEEEtran}
\IEEEoverridecommandlockouts
\usepackage{cite}
\usepackage{amsmath,amssymb,amsfonts}
\usepackage{algorithmic} 
\usepackage{graphicx}
\usepackage{textcomp}
\usepackage{xcolor}
\usepackage{url}

\def\BibTeX{{\rm B\kern-.05em{\sc i\kern-.025em b}\kern-.08em
    T\kern-.1667em\lower.7ex\hbox{E}\kern-.125emX}}
\begin{document}
\makeatletter
\newcommand{\linebreakand}{%
  \end{@IEEEauthorhalign}
  \hfill\mbox{}\par
  \mbox{}\hfill\begin{@IEEEauthorhalign}
}

\makeatother

\title{ A Multiclass Acoustic Dataset and Interactive Tool for Analyzing Drone Signatures in Real-World Environments
}

\author{\IEEEauthorblockN{Mia Y. Wang*}
\IEEEauthorblockA{\textit{Department of Computer Science} \\
\textit{College of Charleston}\\
Charleston, USA \\
wangy5@cofc.edu}
\and

\IEEEauthorblockN{Mackenzie Linn}
\IEEEauthorblockA{\textit{Department of Computer Science} \\
\textit{College of Charleston}\\
Charleston, USA \\
linnmj@g.cofc.edu}

\linebreakand

\IEEEauthorblockN{Andrew P. Berg}
\IEEEauthorblockA{\textit{Department of Computer Science} \\
\textit{College of Charleston}\\
Charleston, USA \\
berga2@g.cofc.edu}

\and

\IEEEauthorblockN{Qian Zhang*\thanks{*Corresponding author.}}
\IEEEauthorblockA{\textit{Department of Engineering} \\
\textit{College of Charleston}\\
Charleston, USA \\
zhangq@cofc.edu}

}

\maketitle

\begin{abstract}
The rapid proliferation of drones across various industries has introduced significant challenges related to privacy, security, and noise pollution. Current drone detection systems, primarily based on visual and radar technologies, face limitations under certain conditions, highlighting the need for effective acoustic-based detection methods. This paper presents a unique and comprehensive dataset of drone acoustic signatures, encompassing 32 different categories differentiated by brand and model. The dataset includes raw audio recordings, spectrogram plots, and Mel-frequency cepstral coefficient (MFCC) plots for each drone. Additionally, we introduce an interactive web application that allows users to explore this dataset by selecting specific drone categories, listening to the associated audio, and viewing the corresponding spectrogram and MFCC plots. This tool aims to facilitate research in drone detection, classification, and acoustic analysis, supporting both technological advancements and educational initiatives. The paper details the dataset creation process, the design and implementation of the web application, and provides experimental results and user feedback. Finally, we discuss potential applications and future work to expand and enhance the project.
\end{abstract}


\section{Introduction}
The rapid proliferation of drones in various industries such as delivery, surveillance, agriculture, and entertainment has introduced significant challenges and opportunities. While drones offer numerous benefits, their widespread use has also raised concerns regarding privacy, security, and noise pollution. Unauthorized drone activity can lead to breaches of privacy and potential security threats, while drones contribute to environmental noise pollution, affecting human health and wildlife. Current drone detection systems primarily rely on visual and radar-based technologies, which face limitations under poor visibility or in cluttered environments. Acoustic-based detection presents a promising complementary approach, but there is a notable lack of comprehensive acoustic datasets encompassing a wide range of drone models and operational conditions. This gap hinders the development of robust detection algorithms and effective noise mitigation strategies. Additionally, there is a need for interactive tools to facilitate research and education in drone acoustics.

This journal paper extends our previous work on drone visualization, which was presented in the 2024 Artificial Intelligence x Humanities, Education, and Art (AIxHeart 2024) Conference \cite{wang2024comprehensive}.
To address these challenges, we present a novel dataset comprising audio recordings, spectrograms, and Mel-frequency cepstral coefficient (MFCC) plots for 32 different drone categories, differentiated by brand and model. Alongside this dataset, we introduce an interactive web application designed to allow users to explore the data intuitively. Users can select specific drone categories, listen to the associated audio recordings, and view the corresponding spectrogram and MFCC plots. This tool aims to enhance research capabilities in drone detection, classification, and acoustic analysis, support noise mitigation efforts, and serve as an educational resource. The publicly available website can be found online \footnote{\url{https://mackenzie-jane.github.io/drone-visualization/}}.

The rest of this paper is organized as follows: Section 2 provides a detailed review of the literature related to drone acoustic detection, classification, and noise pollution, as well as existing datasets and interactive tools. Section 3 describes the dataset, including methods of data collection and the formats of the audio recordings, spectrograms, and MFCC plots. Section 4 discusses the design and implementation of the visualization web application, outlining its user interface and backend architecture. Section 5 presents experimental results, including an analysis of the dataset and user feedback on the web application. In Section 6, we explore potential applications of our dataset and tool, and outline future work to expand and enhance the project. Finally, Section 7 concludes the paper, summarizing our contributions and the impact of our research.

\section{Literature Review}
\label{sec:literature-review}
The rapid growth of drone technologies and acoustic sensing capabilities has sparked a diverse body of research spanning detection techniques, sensing modalities, dataset development, and interactive tools for education and exploration. This section provides a structured review of prior work in seven key areas. We begin by summarizing general audio-based detection methods, followed by vision-based and radar-based approaches, each offering unique benefits and limitations. We then review detection systems based on radio frequency signatures and explore available datasets designed to support classification and benchmarking tasks. Finally, we highlight interactive tools that promote hands-on learning and simulation-based exploration in both formal education and public-facing research platforms.


\subsection{Audio-Based Methods for UAV Detection}
Acoustic sensing offers a unique, low cost, and passive modality for unmanned aerial vehicle (UAV) detection, particularly effective in scenarios where visual or radio frequency based systems may be limited by occlusion, range, or signal interference. When drones operate, their motors and rotors emit characteristic sounds that vary across models, providing an opportunity to capture distinctive audio fingerprints. These acoustic signatures can be used not only for detection but also for identification and classification, especially when represented using time frequency features such as spectrograms and Mel Frequency Cepstral Coefficients (MFCCs)~\cite{wang2021feature, wang2022feature}.

Several studies have leveraged these acoustic characteristics to build UAV detection datasets and evaluate various signal processing pipelines. Wang et al.~\cite{wang2021feature} compared five feature extraction techniques available in the Librosa Python library—MFCCs \cite{mcfee2015librosa}, chroma, Mel spectrograms, spectral contrast, and tonnetz—applied to audio recordings collected from DJI Phantom 4 and EVO 2 Pro drones, along with environmental noise samples. Their analysis revealed that combining multiple acoustic features significantly enhanced the discriminative capacity of the data.

Other researchers have explored real time detection potential and robustness to noise. Jeon et al.~\cite{jeon2017empirical} studied UAV detection within a 150 meter range using Gaussian Mixture Models (GMM), Convolutional Neural Networks (CNN), and Recurrent Neural Networks (RNN). They augmented drone recordings with environmental sounds and evaluated MFCC and Mel spectrogram features. Their results confirmed that acoustic signal fidelity degrades beyond 150 meters, reinforcing the importance of proximity in audio based systems.

Seo et al.~\cite{seo2018drone} used Short Time Fourier Transform (STFT) features with CNNs to evaluate detection accuracy and false alarm rates in open air environments. Their dataset, built from hovering Phantom 3 and Phantom 4 drones, achieved high detection rates exceeding 98\% and a low false alarm rate of 1.28\%, showcasing the utility of normalized time frequency features for clean drone recordings.

To address data scarcity, Al Emad et al.~\cite{al2021audio} generated synthetic UAV audio using Generative Adversarial Networks (GANs). Their hybrid dataset supported both binary classification (drone versus noise) and multi class identification tasks. CNN, RNN, and CRNN models trained on this augmented corpus demonstrated improved generalizability, highlighting the potential of synthetic data for expanding UAV sound libraries.

More recently, Kim et al.~\cite{kim2023self} introduced a self-supervised learning framework to mitigate the limitations of label dependent models. Their approach transforms drone audio into MFCC-based image representations and applies SimCLR, a contrastive learning algorithm, to learn generalized latent features. Without requiring labeled data, their model achieved classification accuracy on par with supervised baselines—reaching a top-1 accuracy of 87.91\%. Notably, the system generalized to unseen drone types, demonstrating strong potential for scalable and adaptive drone detection in real world settings.

Collectively, these studies emphasize the need for diverse, well annotated, and context rich UAV acoustic datasets to support future research in drone detection and sound based classification. Our work builds upon this foundation by providing a comprehensive dataset of 32 UAV classes with corresponding MFCC and Mel spectrogram visualizations, aimed at enabling both analytical exploration and educational applications through an interactive web tool.

\subsection{Vision Based Methods for UAV Detection}
Computer vision has emerged as one of the most widely adopted approaches for UAV detection and classification due to the proliferation of cameras and advances in machine learning based image analysis. Vision based systems rely on optical or infrared imagery to locate and identify drones based on their shape, motion patterns, and appearance. However, they face challenges in low light conditions, visual occlusion, and adverse weather.

Aker and Kalkan\cite{aker2017using} developed an end to end object detection framework using the You Only Look Once (YOLO) architecture, a real time single shot detector built on convolutional neural networks (CNNs). Their system was trained using a dataset composed of bird and drone images embedded in varied backgrounds to simulate real world variability. The trained model demonstrated high precision and recall (both at 0.9), confirming the feasibility of rapid detection from video frames.

Rozantsev et al.\cite{rozantsev2016detecting} proposed motion stabilization techniques to enhance the visual classification of UAVs from moving cameras. By extracting spatio temporal features from image cubes and applying regression based stabilization, their system improved object detection in dynamic scenes. They evaluated boosted tree and CNN based classifiers on custom collected datasets comprising UAVs and aircraft, reporting average precision scores above 0.84 for UAV categories.

Lee et al.\cite{lee2018drone} proposed a two module system that combines a drone detection module and an identification module. The detection module used Haar like features and OpenCV's object detection pipeline, while the identification module applied a simple CNN with two convolutional and two fully connected layers. Their dataset included 7,000 drone and 3,000 non drone images. The overall system achieved 89\% detection accuracy and 91.6\% identification accuracy.

These vision based techniques offer robust performance in clear visual environments and are well suited for medium to long range UAV detection. However, they are less effective under occlusion, poor lighting, or fast drone maneuvers, motivating the need for complementary modalities such as acoustics or radar.

\subsection{Radar Based Methods for UAV Detection}
Radar based detection methods provide an effective alternative for identifying UAVs, especially in scenarios involving poor visibility or long distance operation. These systems detect objects by emitting radio waves and analyzing the reflected signals, offering advantages in range, reliability, and performance under challenging environmental conditions.

Mendis et al.\cite{mendis2016deep} developed a radar detection framework using an S band continuous wave radar coupled with a deep belief network (DBN). They extracted micro Doppler signatures from three different UAV types, including a helicopter, an artificial bird, and a quadcopter. By using spectral correlation functions as inputs, their DBN based classifier achieved over 90\% classification accuracy. Gaussian noise was added for data augmentation, which helped evaluate system performance under various signal conditions.

Kim et al.\cite{kim2016drone} proposed a system using a frequency modulated continuous wave radar along with a pre trained convolutional neural network, specifically GoogleNet. Their dataset included over 66,000 micro Doppler and cadence velocity diagram images collected both indoors in an anechoic chamber and outdoors. By simulating different motor types and observation angles, they assessed model robustness and achieved 94.7\% classification accuracy across varied scenarios.

Radar based approaches demonstrate strong performance in tracking UAV movement across large distances and in detecting small objects under low light or visually obstructed environments. However, these systems typically require more specialized hardware and signal processing expertise, which may limit widespread adoption compared to vision or acoustic based systems.

\subsection{Radio Frequency Based Methods for UAV Detection}
Radio frequency based detection systems leverage the electromagnetic emissions produced by the communication link between UAVs and their controllers. These systems can detect and identify drones based on signal characteristics such as transmission frequency, modulation patterns, and signal strength, making them particularly useful for detecting both the drone and the pilot’s control source.

Ezuma et al.\cite{ezuma2019micro} presented a detection framework that uses radio frequency transmissions between UAVs and their controllers to distinguish UAV related activity from background communication signals. Their approach employed a Bayesian model derived from Markov processes to perform binary detection and multi class classification. Extracted features included spectral entropy, skewness, variance, and kurtosis of the signal, followed by feature selection using neighborhood component analysis. The processed features were input into several machine learning classifiers, including support vector machines and neural networks. Their model achieved over 96\% classification accuracy.

Zhao et al.\cite{zhao2017detection} proposed a detection method based on Gaussian Mixture Models and an adaptive thresholding mechanism to determine UAV activity. They collected data from seven UAV models, extracting signal onset points through statistical analysis of Gaussian distributions. Their method achieved 97\% detection accuracy, demonstrating strong performance in identifying the beginning of UAV control signal activity.

Sazdi\'c Joti\'c et al.\cite{sazdic2022drone} conducted a comprehensive review of radio frequency based UAV detection and classification methods. Their study categorized techniques into classical, advanced, and hybrid engineering approaches, highlighting the strengths and limitations of each in relation to input data types such as MAC addresses, communication protocols, extracted features, and raw I/Q signals. They emphasized that deep learning based methods, particularly those leveraging raw RF signals and time frequency representations, have shown superior performance in recent literature. Furthermore, they discussed key publicly available RF drone datasets, such as the DroneRF and VTI RF datasets, and noted the scarcity of large scale open access RF datasets. Their comparative analysis showed that deep learning models trained on the DroneRF and VTI RF datasets consistently achieved high classification performance. For example, CNN-based models on the DroneRF dataset reported up to 100\% detection accuracy and over 94\% type identification accuracy~\cite{sazdic2022drone}. Similarly, models evaluated on the VTI\_RF\_Dataset achieved over 99\% in drone detection and 97\% in multiple drone identification scenarios. These results underscore the effectiveness of modern deep learning techniques when paired with well-curated RF signal datasets.

Radio frequency based techniques provide the advantage of long range and real time detection capabilities, particularly in open environments where signal propagation is reliable. However, these methods can be challenged by encrypted communication, frequency hopping protocols, and legal restrictions on RF signal monitoring. As such, they are often used in conjunction with other sensing modalities for comprehensive UAV detection solutions.

\subsection{Audio-Derived Visual Methods}

Recent work has explored the transformation of acoustic data into visual formats to leverage the strength of computer vision models in drone classification. Kim et al.\cite{kim2024improving} introduced a novel methodology that converts audio recordings into Mel Frequency Cepstral Coefficient (MFCC) plots, enabling the use of image-based deep learning models. Their dataset included 32 UAV categories, each with 100 five-second audio recordings, from which MFCC visualizations were generated.

The authors trained and compared three vision architectures—EfficientNet, ResNet50, and Vision Transformer—on these MFCC images. Among them, EfficientNet achieved the highest classification accuracy of 96.31\%, followed by ResNet50 at 94.22\%, and Vision Transformer at 73.69\%. These results highlight the promise of combining auditory signals with visual model pipelines and underscore the discriminative power of MFCCs when processed through image-based frameworks.

This audio-to-visual strategy provides a compelling hybrid solution that bridges the strengths of acoustic data and modern vision models. It also opens new opportunities for multi-modal fusion in UAV detection, particularly in settings where raw audio may be harder to interpret directly.



\subsection{Existing UAV Audio Datasets}

Despite promising advancements in UAV detection through acoustic sensing, the progress of audio-based systems has been constrained by the limited availability of high-quality, publicly accessible datasets. Most existing efforts in this domain focus on creating targeted corpora tailored to specific drone models or environmental conditions, which restricts their utility for broader model generalization and benchmarking.

Wang et al.~\cite{wang202415} introduced one of the largest open-access UAV audio datasets to date, containing recordings from 15 different drones—including both small toy models and larger Class I UAVs—totaling 8,120 seconds of annotated audio. The dataset captures diverse operational sounds and was used to train a convolutional neural network for 15-class classification, achieving an average test accuracy of 98.7\% and a test loss of 0.076. These results underscore the dataset’s value for supporting robust, real-world classification systems.

Building on this work, Wang et al.~\cite{Wang2024} further investigated the impact of feature design on UAV classification performance. Specifically, they evaluated various quantities of Mel Frequency Cepstral Coefficients (MFCCs) and determined that using 30 coefficients provided an optimal balance between feature richness and noise resilience. This study also introduced a companion image-based dataset derived from the original audio, consisting of waveform, spectrogram, Mel filter bank, and MFCC plots across 26 UAV categories. With 100 audio samples per category, the dataset supports both visual and audio modality exploration and facilitates the development of multimodal detection systems.

Together, these datasets offer a foundation for standardized evaluation in acoustic-based UAV detection and serve as critical resources for researchers aiming to improve generalization, scalability, and interpretability of drone classification models.

\subsection{Interactive Tools for Research and Education}

Interactive tools for audio-based exploration and simulation are increasingly used to enhance understanding of acoustic principles, foster student engagement, and support open-ended inquiry in both research and classroom settings.

Interactive tools and platforms for exploring acoustic data are crucial for advancing research and education. Projects like BirdVox~\cite{lostanlen2022birdvox}, which provides interactive access to bird sound datasets, illustrate the benefits of such tools. However, similar resources for drone acoustics are notably lacking. This gap hampers the ability of researchers to conduct in-depth analyses and limits the educational potential of these datasets.

Moheit et al.~\cite{moheit2021acoustics} introduced the Acoustics Apps platform, a browser-based e-learning environment that uses high-fidelity simulations powered by COMSOL Server technology. These apps support interactive exploration of complex wave phenomena, musical instrument behavior, and room acoustics without requiring access to physical lab equipment. Designed to be intuitive and device-independent, Acoustics Apps have been successfully used in both high school and university settings to visualize invisible acoustic behaviors and engage students through virtual experiments and self-guided exploration.

In the context of sonic interaction design, Delle Monache et al.~\cite{monache2010toolkit} presented the Sound Design Toolkit (SDT), a modular software environment for real-time, physics-based sound synthesis. SDT includes a library of sound models—such as impacts, friction, and fluid sounds—that can be interactively controlled using sensors or mapped to MIDI/OSC inputs. Developed with education and prototyping in mind, SDT facilitates experiential learning by enabling users to sketch, manipulate, and evaluate sonic feedback in design scenarios. Its taxonomy of everyday sounds also makes it suitable for classroom demonstrations of sound physics and design aesthetics.

Arabasi et al.~\cite{arabasi2018visual} designed a simple, interactive GUI tool in MATLAB that allows students to record and visualize sound waveforms and their corresponding frequency spectra. Primarily used to teach the Fourier transform concept in introductory engineering courses, the tool lets users experiment with different input sounds—including their own voice—and immediately observe how spectral components vary. This low-cost and hands-on approach is particularly effective in demystifying frequency-domain analysis for first-year students.

Tawil and Dahlan~\cite{tawil2021application} examined the role of Interactive Audio Visual (IAV) media in improving creative thinking among science students. Their mixed-methods study found that students who engaged with IAV materials demonstrated significantly higher creative thinking scores compared to those using conventional PowerPoint media. Students also expressed high levels of interest and ease in using the interactive content, noting improved comprehension and increased motivation during tasks involving simulation-based learning. These findings reinforce the educational value of interactive multimedia in facilitating critical and creative thinking skills.

Wang et al. \cite{wang2024comprehensive} developed a web-based drone audio visualization tool that enables users to explore the unique acoustic signatures of drones by listening to recordings and examining their associated spectrogram and MFCC plots. This platform laid the groundwork for the current journal study by providing an initial system for drone sound data visualization. The updated version enhances this foundation with expanded features, improved interactivity, and integration of a broader UAV dataset. The platform presents a browsable interface with drone images and playback controls, designed to make drone acoustics accessible for students, researchers, and hobbyists.

Collectively, these platforms reflect a growing emphasis on accessible, engaging, and interactive resources in acoustics education and sonic research. They demonstrate how interactivity—whether through sound manipulation, simulation, or visualization—can significantly deepen conceptual understanding and foster interdisciplinary learning.

\begin{table*}[ht]
\centering
\caption{UAV Audio Dataset: 32 Classes with Collection Sites}
\label{tab:uav_data_32}
\begin{tabular}{lllrrl}
\hline
\textbf{Manufacture} & \textbf{Model} & \textbf{Drone Type} & \textbf{Number of Files} & \textbf{Duration (sec)} & \textbf{Collection Site} \\
\hline
Self-build & David Tricopter & Outdoor & 100 & 500 & Columbus, IN \\
Self-build & PhenoBee & Outdoor & 100 & 500 & West Lafayette, IN \\
Autel & Evo 2 Pro & Outdoor & 100 & 500 & New Richmond, IN \\
DJI & Avata & Outdoor & 100 & 500 & Charleston, SC \\
DJI & FPV & Outdoor & 100 & 500 & Charleston, SC \\
DJI & Matrice 200 & Outdoor & 100 & 500 & West Lafayette, IN \\
DJI & Matrice 200 V2 & Outdoor & 100 & 500 & New Richmond, IN \\
DJI & Matrice 600p & Outdoor & 100 & 500 & New Richmond, IN \\
DJI & Mavic Air 2 & Outdoor & 100 & 500 & New Richmond, IN \\
DJI & Mavic Mini 1 & Outdoor & 100 & 500 & New Richmond, IN \\
DJI & Mini 2 & Outdoor & 100 & 500 & New Richmond, IN \\
DJI & Mini 3 & Outdoor & 100 & 500 & Charleston, SC \\
DJI & Mini 3 Pro & Outdoor & 100 & 500 & Charleston, SC \\
DJI & Mavic 2 Pro & Outdoor & 100 & 500 & New Richmond, IN \\
DJI & Neo & Outdoor & 100 & 500 & Charleston, SC\\
DJI & Mavic 2s & Outdoor & 100 & 500 & New Richmond, IN \\
DJI & Phantom 2 & Outdoor & 100 & 500 & New Richmond, IN \\
DJI & Phantom 4 & Outdoor & 100 & 500 & New Richmond, IN \\
DJI & Tello & Indoor & 100 & 500 & Charleston, SC \\
DJI & RoboMaster TT Tello & Indoor & 100 & 500 & New Richmond, IN \\
Hasakee & Q11 & Indoor & 100 & 500 & West Lafayette, IN \\
Holystone & HS210 & Indoor & 100 & 500 & Charleston, SC \\
Hover & X1 & Outdoor & 100 & 500 & Charleston, SC \\
Syma & X5SW & Indoor & 100 & 500 & West Lafayette, IN \\
Syma & X5UW & Indoor & 100 & 500 & West Lafayette, IN \\
Syma & X8SW & Indoor & 100 & 500 & West Lafayette, IN \\
Syma & X20 & Indoor & 100 & 500 & West Lafayette, IN \\
Syma & X20P & Indoor & 100 & 500 & West Lafayette, IN \\
Syma & X26 & Indoor & 100 & 500 & West Lafayette, IN \\
Swellpro & Splash 3 plus & Outdoor & 100 & 500 & New Richmond, IN \\
Yuneec & Typhoon H Plus & Outdoor & 100 & 500 & New Richmond, IN \\
UDI RC & U46 & Outdoor & 100 & 500 & West Lafayette, IN \\
\hline
 & & Total & 3,200 & 16,000\\
\hline
\end{tabular}
\end{table*}

\section{Methodology}
\subsection{Data Collection}
The drone data collection is an ongoingt multi-year effort aimed at building a large-scale, diverse dataset of UAV acoustic signatures~\cite{wang2022large, wang202415, wang2024comprehensive, Wang2024}. As of 2025, the dataset comprises 3,200 audio recordings captured from 32 distinct unmanned aerial vehicles (UAVs), totaling 16,000 seconds of raw flight audio. Each UAV contributed 100 five-second audio clips. These recordings span a wide range of drone types and environments and serve as the foundation for acoustic analysis, feature extraction, and educational visualization.

\textbf{Drone Overview:} The collection includes 28 quadcopters, one tricopter, two hexacopters, and one tail-sitter UAV. The majority feature standard X-frame quadrotor configurations. Drone platforms include commercial and consumer models from DJI, Autel, Syma, Yuneec, UDI, Hasakee, Holystone, and Hover, as well as two custom-built designs. Notable entries include the \textit{David Tricopter}, a custom-built tricopter with a 34-inch diameter and AfroFlight Naze32 flight controller, and \textit{PhenoBee}, a large-scale hexacopter weighing 23 kg, designed by Ziling Chen and built on the Ardupilot Cube Orange platform.

\textbf{Recording Sites:} Audio recordings were collected in diverse indoor and outdoor environments across three U.S. locations: West Lafayette, Indiana; New Richmond, Indiana; and Charleston, South Carolina. Indoor data from Indiana were acquired in university laboratories, while outdoor recordings were made on a private farm in New Richmond. Charleston-based data were collected in the College of Charleston's Drone Lab at the Harbor Walk Campus (indoor) and from the rooftop of the South Carolina Aquarium parking garage (outdoor). Recordings captured natural environmental noise such as wind, birdsong, and traffic, contributing to a realistic audio corpus.

\textbf{Recording Equipment:} From 2021 to 2023, data were recorded using a MacBook Air (1.1GHz quad-core Intel Core i5, 8GB RAM) with the system's internal microphone. Beginning in 2024, recordings were made with an updated MacBook Air featuring an Apple M3 chip and 16GB of memory. No external microphones or post-processing techniques were used, preserving the raw acoustic characteristics of each drone.

This dataset underpins the visual and analytical tools presented in this study, including the expanded web-based interface for exploring drone-specific acoustic features such as MFCCs and spectrograms.

\subsection{Visualization Dataset Creation}

The project's implementation uses Librosa \cite{mcfee2015librosa} to compute the Mel Frequency Cepstral Coefficient (MFCC). Our number of mfcc were set to 20 (n-mfcc, FFT window size to 2048 (n-fft), overlap between frames 512 (hop length), and the number of mels to 128 (n-mels). The mathematical descriptions below reflect what is abstracted in the Librosa package.

Extracting MFCCs from an audio dataset involves several steps. The process begins with digital audio files (i.e. .wav, .mp3, .ogg, etc), representing the raw audio signal. The audio is segmented into short overlapping windows ranging from 20 to 40 milliseconds. To reduce signal noise, the Hanning window function is applied, it is mathematically given as used by Harris \cite{harris_hanning_1978}:

\[
w[n] = \frac{1}{2}[1-cos(\frac{2 \pi n}{N})]  , ~ \text{for} ~ 0 \le n \le N-1.
\]

where $w[n]$ is the Hanning window function and $N$ is the total number of windows to be computed. Note that we are using zeroth indexing in the function above. This improves the accuracy of the following feature representations, by smoothing the signal with the $1-cos(\frac{2 \pi n}{N})$ term.

Next, the Short-Time Fourier Transform is applied (STFT). Specifically, the continuous-time STFT is applied. It can mathematically be given as:

$x(t, \omega)$:
\[
\text{STFT}_x(t, \omega) = \int_{-\infty}^{\infty} x(\tau)\, w(\tau - t)\, e^{-j \omega \tau} \, d\tau
\]

The STFT is computed for each windowed frame.Using the $w(\tau - t)$ time-centered windowing function segments the raw signal $x(\tau)$ onto the STFT's sinusoidal basis function $e^{-j \omega \tau}$. the $\text{STFT}_x(t, \omega)$. If computation stopped at this step, the plot would be a spectrogram.

Next the mel-scale is applied. Which was developed to mimic the human perception of hearing. It isn't critical to model performance, but is consistent in related literature \cite{wang2021feature}\cite{seo2018drone}.
Mathematically the mel-scale can be written as:

\[
m(f) = 2595 \cdot \log_{10}\left(1 + \frac{f}{700}\right)
\]

The mel-scale transform converts frequencies from Hertz (Hz) to those in the mel-scale (mels). If computation stopped, the plot would be a mel-spectrogram

From here the approach uses triangular filter banks to calculate the relative amplitude of the frequencies. Shown is the piecewise definition as used in Hang Xu et al. \cite{Haung2001Language}:
\[
H_m[k] =
\begin{cases}
  0 & k < f[m-1] \\
  \frac{k - f[m-1]}{f[m] - f[m-1]} & f[m-1] \le k \le f[m] \\
  \frac{f[m+1] - k}{f[m+1] - f[m]} & f[m] \le k \le f[m+1] \\
  0 & k > f[m+1]
\end{cases}
\]

$m$ is the mel filter index and $k$ is the index for the frequency bin. If computation stopped here, the plot would be considered a mel-filterbank. 

In the final step, the processed signal is log-scaled and then passed through the Discrete Cosine Transform (DCT) of the mel log signal. Specifically the DCT-II formalization, which is standard for audio processing. Mathematically it is defined as:

\[
    X_k =
 \sum_{n=0}^{N-1} x_n \cos \left[\, \tfrac{\,\pi\,}{N} \left( n + \tfrac{1}{2} \right) k \, \right]
 \qquad \text{ for } ~ k = 0,\ \dots\ N-1 ~.
\]

The formulation of the augmented base cosine function allows the cosine function
$\cos \left[\, \tfrac{\,\pi\,}{N} \left( n + \tfrac{1}{2} \right) k \, \right]$ to give unique information for each frequency component, lending itself for an efficient orthogonal representation without waste.

The DCT transform converts $N$ time/spatial samples into $N$ frequency coefficients: $[x_0, x_1, \ldots, x_{N-1}] \rightarrow [X_0, X_1, \ldots, X_{N-1}]$.

After applying the DCT, the MFCC is computationally complete, resulting in a compact and rich representation of an original audio signal.


We have generated a total of 3,200 MFCC plots extracted from audio recordings across 32 categories, with each category containing 100 audio files. These MFCC plots serve as feature-rich representations of the acoustic characteristics captured from the audio data; essential for further analysis and classification tasks, reflecting the unique acoustic signatures of various UAV drone audio recordings.

\subsection{Web Application Development}
The Drone Audio Visualization Tool is an interactive web application designed to enhance the exploration and analysis of a drone audio dataset. Its user interface (UI) is designed so that users can intuitively navigate the application, explore the dataset, and gain meaningful insight into drone audio patterns. The publicly available website can be accessed at: \url{https://mackenzie-jane.github.io/drone-visualization/#/}. 

Users begin by opening the homepage, which provides an overview of the project and a selection of all 32 drone images. This serves as a visual entry point into the dataset and facilitates quick orientation between pages. Figure \ref{fig: home} illustrates the layout of the home page, which shows the drone images and basic information. 

\begin{figure}[htp]
  \centering    
  \includegraphics[width=0.5\textwidth]{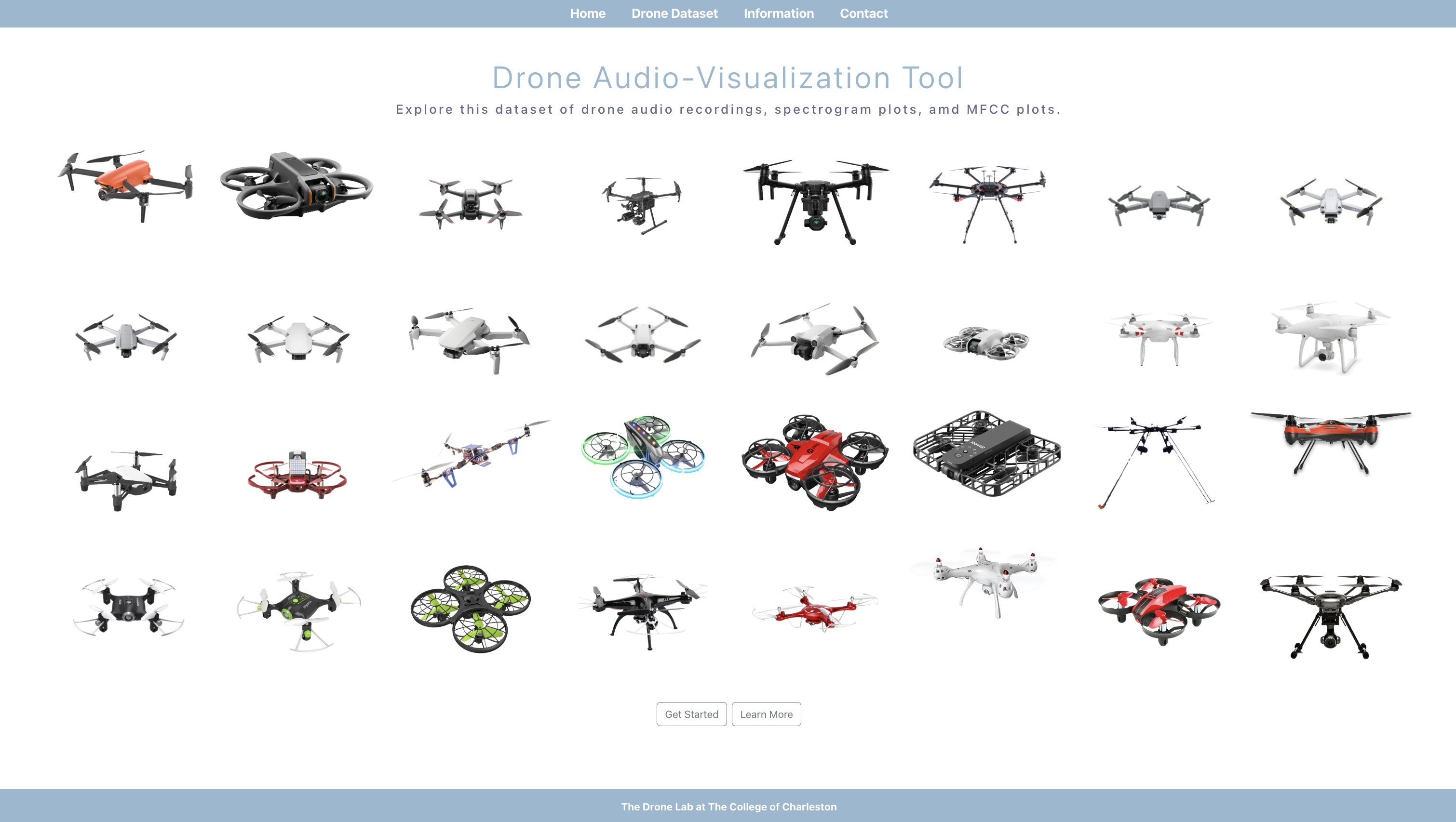}
  \caption{Audio Visualization Tool Web Application Home Page}
  \label{fig: home}
\end{figure}

From the homepage, users can navigate to the drone dataset page, which features a responsive grid layout of 32 cards, one for each drone in the dataset. Each card includes the drone’s name and image, enabling quick identification and selection. This intuitive and visually appealing layout helps users quickly identify and select their drone of interest. 

Upon selecting a drone, users are redirected to the drone detail page. This page presents a cohesive view of the attributes of the selected drone, including its name, image, an audio recording sample, and two visualizations: a Mel-Frequency Cepstral Coefficient (MFCC) plot and a spectrogram plot. These visualizations are generated at a randomly selected frame, offering a snapshot of the drone’s acoustic signature. Figure \ref{fig: drone} shows an example of the drone information and plots. The layout ensures that the visualizations, drone image, and audio information are presented in a cohesive way. This enables users to simultaneously see and hear key characteristics of each drone, supporting both qualitative and quantitative analysis of drone sound profiles.

\begin{figure}[htp]
  \centering    
  \includegraphics[width=0.5\textwidth]{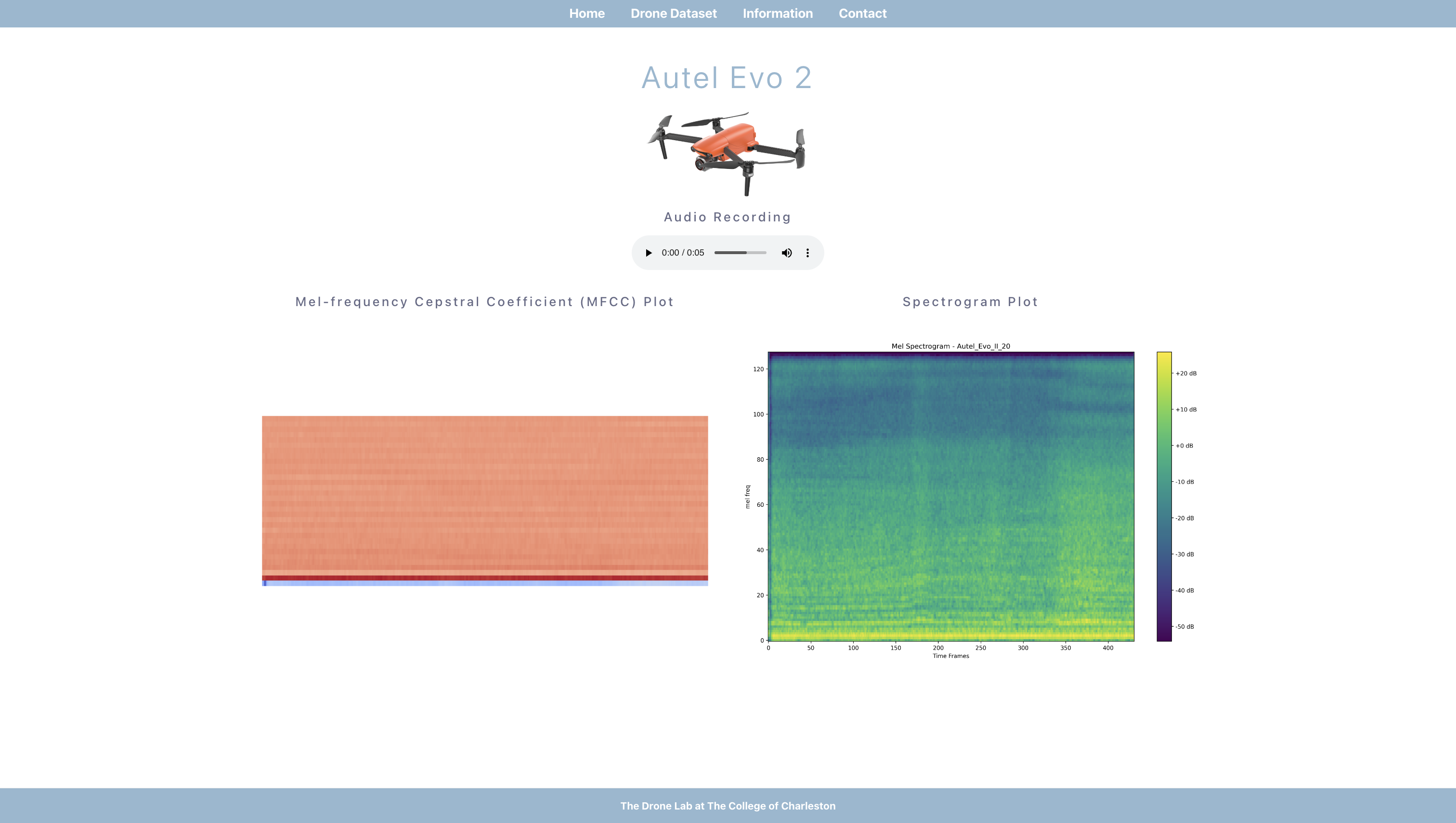}
  \caption{Audio Visualization Tool Drone Display Page}
  \label{fig: drone}
\end{figure}

This website is built using the React framework to structure and render dynamic components for each drone. In addition, CSS is used to style the interface and ensure responsiveness across devices. JavaScript enables interactivity, such as page transitions and dynamic content loading. Media files, drone images, MFCC plots, spectrograms, and audio files are stored in organized subdirectories within the public folder. A structured JSON file maps drone identifiers to their corresponding media and metadata. This architecture supports efficient and dynamic content loading.

The dataset includes audio recordings, spectrogram images, and MFCC plots for various drones at specific frames ranging from 0 to 100. A Python script was used to randomly select frame numbers between 0 and 100 for each drone. These random frame indices are used to extract specific audio segments and the corresponding MFCC and spectrogram plots for each drone. This approach ensures an unbiased and varied sampling, which is useful for identifying distinguishing acoustic features across different drones.

Using static file storage and dynamic content loading, the web application provides an efficient and user-friendly platform for drone audio visualization. This website improves accessibility to the dataset and supports further research in drone classification and analysis. The source code for the Drone Audio Visualization Tool is publicly available on GitHub at: \url{https://github.com/mackenzie-jane/drone-visualization}.

\section{Conclusion}

The rapid growth of drone usage across various industries has introduced significant challenges, such as privacy, security, and noise pollution, that current visual and radar-based detection systems struggle to address. Our research tackles these issues by introducing a comprehensive dataset of drone acoustic signatures, covering 32 categories by brand and model. This dataset includes raw audio recordings, spectrogram plots, and Mel-Frequency Cepstral Coefficient (MFCC) plots, providing a robust foundation for acoustic-based drone detection and classification.

In addition to the dataset, we developed an interactive web application that allows users to explore drone acoustic signatures. Users can select specific drone categories, listen to the corresponding audio, and view the associated spectrogram and MFCC plots. This tool supports both technological advancements in drone detection and educational initiatives, making it accessible to a broad audience.

We detailed the dataset creation process, the design and implementation of the web application, and presented experimental results and user feedback. The positive reception and high accuracy rates achieved in our experiments demonstrate the potential of acoustic-based methods for drone detection.

Looking ahead, there are numerous opportunities for expanding and enhancing this project. Future work could involve adding more drone models, refining the web application’s features, and exploring advanced machine learning techniques to boost detection accuracy. We also plan to compare the performance of our audio dataset with image datasets by training various deep learning models on both to determine which approach performs best. By continuing to develop and refine these tools, we aim to make meaningful contributions to addressing the challenges related to drone proliferation in terms of privacy, security, and noise pollution.



\bibliographystyle{IEEEtran}
\bibliography{reference}

\begin{thebibliography}{10}
\providecommand{\url}[1]{#1}
\csname url@samestyle\endcsname
\providecommand{\newblock}{\relax}
\providecommand{\bibinfo}[2]{#2}
\providecommand{\BIBentrySTDinterwordspacing}{\spaceskip=0pt\relax}
\providecommand{\BIBentryALTinterwordstretchfactor}{4}
\providecommand{\BIBentryALTinterwordspacing}{\spaceskip=\fontdimen2\font plus
\BIBentryALTinterwordstretchfactor\fontdimen3\font minus \fontdimen4\font\relax}
\providecommand{\BIBforeignlanguage}[2]{{%
\expandafter\ifx\csname l@#1\endcsname\relax
\typeout{** WARNING: IEEEtran.bst: No hyphenation pattern has been}%
\typeout{** loaded for the language `#1'. Using the pattern for}%
\typeout{** the default language instead.}%
\else
\language=\csname l@#1\endcsname
\fi
#2}}
\providecommand{\BIBdecl}{\relax}
\BIBdecl

\bibitem{wang2024comprehensive}
M.~Y. Wang, D.~C. Ramirez, E.~Noonan, M.~Linn, and Q.~Zhang, ``A comprehensive dataset and visualization tool for drone acoustic signatures,'' in \emph{2024 Artificial Intelligence x Humanities, Education, and Art (AIxHEART)}.\hskip 1em plus 0.5em minus 0.4em\relax IEEE, 2024, pp. 13--17.

\bibitem{wang2021feature}
Y.~Wang, F.~E. Fagian, K.~E. Ho, and E.~T. Matson, ``A feature engineering focused system for acoustic uav detection,'' in \emph{2021 Fifth IEEE International Conference on Robotic Computing (IRC)}.\hskip 1em plus 0.5em minus 0.4em\relax IEEE, 2021, pp. 125--130.

\bibitem{wang2022feature}
Y.~Wang, F.~E. Fagiani, K.~E. Ho, and E.~T. Matson, ``A feature engineering focused system for acoustic uav payload detection.'' in \emph{ICAART (3)}, 2022, pp. 470--475.

\bibitem{mcfee2015librosa}
B.~McFee, C.~Raffel, D.~Liang, D.~P. Ellis, M.~McVicar, E.~Battenberg, and O.~Nieto, ``librosa: Audio and music signal analysis in python,'' in \emph{Proceedings of the 14th python in science conference}, vol.~8, no.~14.\hskip 1em plus 0.5em minus 0.4em\relax Citeseer, 2015, pp. 18--25.

\bibitem{jeon2017empirical}
S.~Jeon, J.-W. Shin, Y.-J. Lee, W.-H. Kim, Y.~Kwon, and H.-Y. Yang, ``Empirical study of drone sound detection in real-life environment with deep neural networks,'' in \emph{2017 25th European Signal Processing Conference (EUSIPCO)}.\hskip 1em plus 0.5em minus 0.4em\relax IEEE, 2017, pp. 1858--1862.

\bibitem{seo2018drone}
Y.~Seo, B.~Jang, and S.~Im, ``Drone detection using convolutional neural networks with acoustic stft features,'' in \emph{2018 15th IEEE International Conference on Advanced Video and Signal Based Surveillance (AVSS)}.\hskip 1em plus 0.5em minus 0.4em\relax IEEE, 2018, pp. 1--6.

\bibitem{al2021audio}
S.~Al-Emadi, A.~Al-Ali, and A.~Al-Ali, ``Audio-based drone detection and identification using deep learning techniques with dataset enhancement through generative adversarial networks,'' vol.~21, no.~15.\hskip 1em plus 0.5em minus 0.4em\relax Multidisciplinary Digital Publishing Institute, 2021, p. 4953.

\bibitem{kim2023self}
J.~Kim, M.~Y. Wang, and E.~T. Matson, ``Self-supervised drone detection using acoustic data,'' in \emph{2023 Seventh IEEE International Conference on Robotic Computing (IRC)}.\hskip 1em plus 0.5em minus 0.4em\relax IEEE, 2023, pp. 67--70.

\bibitem{aker2017using}
C.~Aker and S.~Kalkan, ``Using deep networks for drone detection,'' in \emph{2017 14th IEEE International Conference on Advanced Video and Signal Based Surveillance (AVSS)}.\hskip 1em plus 0.5em minus 0.4em\relax IEEE, 2017, pp. 1--6.

\bibitem{rozantsev2016detecting}
A.~Rozantsev, V.~Lepetit, and P.~Fua, ``Detecting flying objects using a single moving camera,'' vol.~39, no.~5.\hskip 1em plus 0.5em minus 0.4em\relax IEEE, 2016, pp. 879--892.

\bibitem{lee2018drone}
D.~Lee, W.~G. La, and H.~Kim, ``Drone detection and identification system using artificial intelligence,'' in \emph{2018 International Conference on Information and Communication Technology Convergence (ICTC)}.\hskip 1em plus 0.5em minus 0.4em\relax IEEE, 2018, pp. 1131--1133.

\bibitem{mendis2016deep}
G.~J. Mendis, T.~Randeny, J.~Wei, and A.~Madanayake, ``Deep learning based doppler radar for micro uas detection and classification,'' in \emph{MILCOM 2016-2016 IEEE Military Communications Conference}.\hskip 1em plus 0.5em minus 0.4em\relax IEEE, 2016, pp. 924--929.

\bibitem{kim2016drone}
B.~K. Kim, H.-S. Kang, and S.-O. Park, ``Drone classification using convolutional neural networks with merged doppler images,'' vol.~14, no.~1.\hskip 1em plus 0.5em minus 0.4em\relax IEEE, 2016, pp. 38--42.

\bibitem{ezuma2019micro}
M.~Ezuma, F.~Erden, C.~K. Anjinappa, O.~Ozdemir, and I.~Guvenc, ``Micro-uav detection and classification from rf fingerprints using machine learning techniques,'' in \emph{2019 IEEE Aerospace Conference}.\hskip 1em plus 0.5em minus 0.4em\relax IEEE, 2019, pp. 1--13.

\bibitem{zhao2017detection}
C.~Zhao, M.~Shi, Z.~Cai, and C.~Chen, ``Detection of unmanned aerial vehicle signal based on gaussian mixture model,'' in \emph{2017 12th International Conference on Computer Science and Education (ICCSE)}.\hskip 1em plus 0.5em minus 0.4em\relax IEEE, 2017, pp. 289--293.

\bibitem{sazdic2022drone}
B.~Sazdi{\'c}-Joti{\'c}, B.~Bond{\v{z}}uli{\'c}, I.~Pokrajac, J.~Baj{\v{c}}eti{\'c}, and M.~Mohammed, ``Drone classification based on radio frequency: techniques, datasets, and challenges,'' in \emph{Conference papers, 10th International Scientific Conference on Defensive Technologies (OTEH 2022)}, 2022.

\bibitem{kim2024improving}
J.~Kim, Q.~Zhang, E.~T. Matson, and M.~Y. Wang, ``Improving drone classification with audio-derived visual features: A vision model comparison,'' in \emph{2024 Eighth IEEE International Conference on Robotic Computing (IRC)}.\hskip 1em plus 0.5em minus 0.4em\relax IEEE, 2024, pp. 41--45.

\bibitem{wang202415}
M.~Y. Wang, Z.~Chu, I.~Ku, E.~Cho~Smith, and E.~T. Matson, ``A 15-category audio dataset for drones and an audio-based uav classification using machine learning,'' \emph{International Journal of Semantic Computing}, pp. 1--16, 2024.

\bibitem{Wang2024}
M.~Wang, Z.~Chu, C.~Entzminger, Y.~Ding, and Q.~Zhang, ``Visualization and interpretation of mel-frequency cepstral coefficients for uav drone audio data,'' in \emph{Proceedings of the 13th International Conference on Data Science, Technology and Applications}, 2024, pp. 528--534.

\bibitem{lostanlen2022birdvox}
V.~Lostanlen, A.~Cramer, J.~Salamon, A.~Farnsworth, B.~M. Van~Doren, S.~Kelling, and J.~P. Bello, ``Birdvox: Machine listening for bird migration monitoring,'' \emph{bioRxiv}, pp. 2022--05, 2022.

\bibitem{moheit2021acoustics}
L.~Moheit, J.~D. Schmid, J.~M. Schmid, M.~Eser, and S.~Marburg, ``Acoustics apps: Interactive simulations for digital teaching and learning of acoustics,'' \emph{The Journal of the Acoustical Society of America}, vol. 149, no.~2, pp. 1175--1182, 2021.

\bibitem{monache2010toolkit}
S.~D. Monache, P.~Polotti, and D.~Rocchesso, ``A toolkit for explorations in sonic interaction design,'' in \emph{Proceedings of the 5th audio mostly conference: a conference on interaction with sound}, 2010, pp. 1--7.

\bibitem{arabasi2018visual}
S.~Arabasi, H.~Al-Taani, and D.~{\"U}. Kapanadze, ``A visual and interactive learning tool: frequency content of sound waves,'' in \emph{EDULEARN18 Proceedings}.\hskip 1em plus 0.5em minus 0.4em\relax IATED, 2018, pp. 10\,719--10\,724.

\bibitem{tawil2021application}
M.~Tawil and A.~Dahlan, ``Application of interactive audio visual media to improve students’ creative thinking skill,'' in \emph{Journal of Physics: Conference Series}, vol. 1752, no.~1.\hskip 1em plus 0.5em minus 0.4em\relax IOP Publishing, 2021, p. 012076.

\bibitem{wang2022large}
Y.~Wang, Z.~Chu, I.~Ku, E.~C. Smith, and E.~T. Matson, ``A large-scale uav audio dataset and audio-based uav classification using cnn,'' in \emph{accpected to publish in 2022 Sixth IEEE International Conference on Robotic Computing (IRC)}.\hskip 1em plus 0.5em minus 0.4em\relax IEEE, 2022.

\bibitem{harris_hanning_1978}
Harris, ``On the use of the windows fro harmonic analysis with the discrete fourier transform,'' IEEE, pp. 60--61, 1978.

\bibitem{Haung2001Language}
A.~Huang and Hon, \emph{Spoken Language Processing: A guide to Theory, Algorithm, and System Development}.\hskip 1em plus 0.5em minus 0.4em\relax Prentic Hall, 2001.

\end{thebibliography}

\end{document}